\documentclass[twocolumn,prb,superscriptaddress,showpacs]{revtex4}
\usepackage{color}
\usepackage{graphicx}
\usepackage{amssymb} 

\newcommand{\surf} {\mbox{\scriptsize surf}}
\newcommand{\mol} {\mbox{\scriptsize mol}}
\newcommand{\coupling} {\mbox{\scriptsize coupling}}

\begin{document}

\title{Many-body electronic structure and Kondo properties of cobalt-porphyrin molecules.}

\author{Luis~G.~G.~V. Dias da Silva}
\altaffiliation{These authors contributed equally to the work.}
\affiliation{Materials Science and Technology Division,
Oak Ridge National Laboratory, Oak Ridge, Tennessee 37831}
\affiliation{Department of Physics
and Astronomy, University of Tennessee, Knoxville, Tennessee
37996}
\email{diasdasilval@ornl.gov}

\author{Murilo L. Tiago}
\altaffiliation{These authors contributed equally to the work.}
\affiliation{Materials Science and Technology Division, Oak
Ridge National Laboratory, Oak Ridge, Tennessee 37831}

\author{Sergio E. Ulloa}
\affiliation{Department of Physics
and Astronomy, Nanoscale and Quantum Phenomena Institute, Ohio
University, Athens, Ohio 45701-2979}

\author{Fernando A. Reboredo}
\affiliation{Materials Science and Technology Division,
Oak Ridge National Laboratory, Oak Ridge, Tennessee 37831}

\author{Elbio Dagotto}
\affiliation{Materials Science and Technology Division, Oak Ridge
National Laboratory, Oak Ridge, Tennessee 37831}
\affiliation{Department of Physics and Astronomy, University of
Tennessee, Knoxville, Tennessee 37996}

\date{\today}

\begin{abstract}
We use a combination  of first principles many-body methods and
the   numerical  renormalization-group  technique to  study
the Kondo regime of cobalt-porphyrin compounds adsorbed on a
Cu(111) surface. We find the  Kondo temperature  to be  highly
sensitive to  both molecule charging and distance to the
surface, which can explain the variations observed in recent
scanning  tunneling spectroscopy  measurements. We discuss  the
importance    of  many-body effects   in   the molecular
electronic structure controlling this phenomenon and suggest
scenarios where enhanced temperatures can be achieved in
experiments.
\end{abstract}







\maketitle

\section{Introduction}
\label{sec:Introduction}

Predicting the electronic properties of  magnetic
nanostructures is  a challenging task  as magnetism  often
manifests itself  as a delicate interplay   between   spin
ordering surface chemistry and  finite size effects.
\cite{reboredo06,TiagoZASC06}  Nanoscale magnetism is   at the
heart of  several proposals for  molecular spintronic devices,
highlighting the need for a detailed understanding of
spin-related phenomena in these systems.   One important and
especially subtle example of this behavior is  the  Kondo
effect, arising  from many-body spin interactions between a
magnetic impurity and electrons in  a metallic
system.\cite{HewsonBook}

At low temperatures, the  Kondo screening of the impurity's
magnetic moment creates a sharp resonance at  the Fermi  energy in
the local density of states at/near the magnetic impurity,
producing a  zero-bias anomalous signal in transport experiments
that probe the local structure. Scanning tunneling  microscopy
(STM) has proved  to be an outstanding tool to probe signatures of
the Kondo effect in magnetic adatoms on metallic surfaces.
\cite{MadhavanCJCW98,LiSBD98,WahlDSVWK04,OtteTBLBLHH08} More
recently, Kondo-like signals have been reported in magnetic
molecules adsorbed on metallic substrates,
\cite{ZhaoLCXWPWXYHZ05,IancuDH06,GaoJHCD07,FuJCMW07,IancuDH06_1}
stimulating          significant        advances    on     the
theory side.\cite{LinCJ05,HuangC06,LeungRSSBCM06,ChiappeL06,Lin:156102:2006,HuangC08,KosekiMTTN08,Roura-BasBL09,Aguiar-Hualde:155415:2009}
For  both single magnetic  atoms and  magnetic  molecules on
metallic surfaces,  the low-bias  conductance  measured    by STM
acquires    a characteristic Fano lineshape,  from which the Kondo
temperature $T_K$ is    usually      inferred       from fittings
to    well-known equations.\cite{UjsaghyKSZ00,PlihalG01} Such
expressions, however, do not  offer   estimates   for $T_K$,
and   rather treat  it as    a phenomenological fitting parameter.
In fact, as we discuss   below, quantitative estimates of $T_K$
are not trivial to obtain in these systems with current
theoretical approaches.

Here,  we  address this issue  by  employing  a combination of
complementary numerical methods to calculate the Kondo
temperature and the density of states  in the  Kondo regime for
complex magnetic molecules using state-of-the-art theories and
a   minimum of information from experimental data. Our approach
relies on  a combination   of two powerful techniques: the
first-principles GW method,\cite{HedinL69,AulburJW00,TiagoC06}
one of the most advanced tools to describe the electronic
structure of  real materials; and the numerical
renormalization-group (NRG),\cite{BullaCP08} widely regarded as
the most reliable numerical method  to study Kondo
correlations.  Our main goals are (i) to identify the relevant
microscopic model and obtain realistic microscopic parameters
from the GW  molecular electronic structure and (ii) to
describe the  Kondo regime in a nonperturbative fashion, with
no additional assumptions on the shape of the  Kondo resonance.

The strength of this merging lies on  the complementarity of
these techniques.  On the one  hand, \textit{ab initio} methods
face formidable challenges in the   Kondo regime since the
strongly-correlated Kondo singlet state  emerges at energy
scales much  smaller than the typical atomic or orbital
energies and is beyond  typical formulations  of exchange and
correlation effects.   Furthermore, the many-body  Kondo state
is not well captured by perturbative approaches, which
sometimes yield infrared  divergencies at    energy  scales of
order of    $k_B T_K$.\cite{HewsonBook}  On the other hand,
NRG, which fully describes the Kondo regime, relies on quantum
impurity models to capture the microscopic low-energy features
of the full system.  In the case of organic complexes
containing magnetic atoms, the presence of molecular states,
additional degrees of freedom and charging effects further
complicates the task of identifying the relevant low-energy
microscopic model leading to the Kondo effect. In fact, the
comparatively high Kondo temperatures ($\gtrsim 100K$) reported
in STM measurements involving porphyrin compounds with a
magnetic atom
\cite{ZhaoLCXWPWXYHZ05,IancuDH06,IancuDH06_1,GaoJHCD07,FuJCMW07}
indicate that extra, molecule- and surface-specific features
play an active role in modifying the Kondo state. In all, such
complexity has hindered attempts to \textit{predict} Kondo
temperature values from first-principle calculations, although
recent work in nanocontacts is most promising.
\cite{Jacob:arXiv:0903.1274:2009}

The presentation is organized as follows: in
Sec.~\ref{sec:Results_GW} we present GW results for the
electronic structure of the molecule. Based on these results,
we propose a realistic microscopic model in
Sec.~\ref{sec:Model} and NRG results for the Kondo temperature
and density of states are presented in
Sec.~\ref{sec:Results_Kondo}. We summarize our findings in
Sec.~\ref{sec:Summary}.

\section{Many-body Electronic Structure}
\label{sec:Results_GW}

In  the   present  work,    we     study  TBrPP-Co   [5, 10,
15, 20-Tetrakis-(4-bromophenyl)-porphyrin-Co, stoichiometry
CoBr$_4$N$_4$C$_{44}$H$_{24}$], which is composed by a
porphyrin ring with four bromophenyl groups at the end parts
and a single cobalt atom at its
center.\cite{IancuDH06_1,IancuDH06,LiYEHKKFX08} The
free-standing TBrPP-Co has $S_4$ symmetry. The anisotropic
environment in the central  cation creates  a four-fold
molecular field which splits the cation's  $3d$ orbitals   into
three nondegenerate   levels    (with $d_{3z^2-1}$,
$d_{x^2-y^2}$,   and  $d_{xy}$  symmetries)    and a degenerate
pair ($d_{xz}$,$d_{yz}$).   The  molecule  adsorbed on a
Cu(111) surface undergoes a planar-saddle deformation that
reduces symmetry even more and  breaks the remaining
degeneracy. \cite{IancuDH06_1} In the following, we restrict
our analysis to the planar  configuration and provide heuristic
arguments for the observed reduction of $T_K$ in the saddle
conformation seen in experiments.

\begin{figure}[tbp]
\includegraphics[clip,width=1\columnwidth]{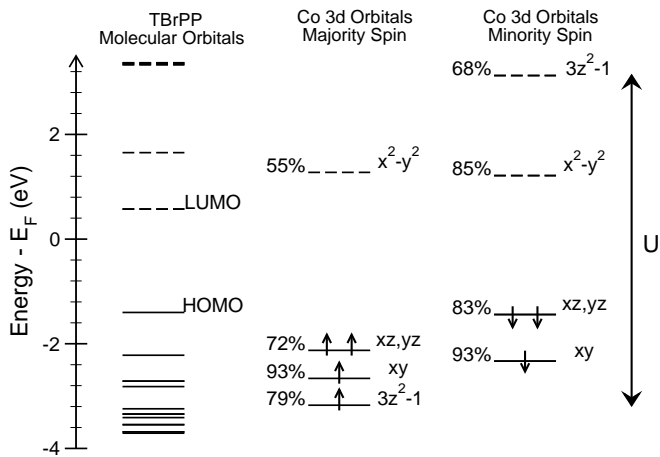}
\caption{Energy of electronic orbitals of TBrPP-Co
in gas phase,
  obtained  from first-principles   GW  calculations.  The  calculated
  Coulomb  energy cost $U$ for  double occupation  of the $d_{3z^2-1}$
  orbital is also indicated.  Orbitals in the middle and right columns
  have projection on Co $3d$ atomic orbitals indicated.  The `` HOMO''
  and ``LUMO'' labels refer,  respectively, to occupied and unoccupied
  orbitals   in the  non-magnetic  TBrPP.    The  Fermi  energy  $E_F$
  indicated  is  an estimate of  the actual Fermi
  energy of adsorbed TBrPP-Co on copper
  (see text).
}
\label{Fig:levels}
\end{figure}

In a preliminary step, we use DFT\cite{Martin} to determine the
electronic structure of TBrPP-Co in gas phase. We use
norm-conserving pseudopotentials\cite{TroullierM91} for the
interaction between valence and core electrons, including
scalar relativistic effects. The exchange-correlation
functional is calculated using the Perdew-Burke-Ernzerhof (PBE)
parametrization.\cite{PerdewBE96} Calculations are done using
the PARSEC code. \cite{ChelikowskyTS94,parsec} We solve
self-consistently the DFT equations in real space, on a regular
grid with grid spacing $0.25~a.u.$, where $1~a.u. = 0.529$~\AA.
Electron wavefunctions are set to zero outside a domain
enclosing the molecule. The distance between the domain
boundary and any atom is at least $4~a.u.$, thus removing any
unwanted effects arising from the scattering of electrons off
the boundary. The DFT results are compatible with previous
calculations on TBrPP-Co\cite{PereraKIDUMH} and similar
molecules.\cite{LiaoS02}

In order to obtain accurate energies of electronic orbitals, we
use the GW method.\cite{AulburJW00,HedinL69,TiagoC06} This
approach is advantageous because it provides very accurate
relative positions of localized $d$ orbitals \emph{and} of more
extended $sp$ orbitals\cite{AulburJW00,KotaniSF07}. More
specifically, most existing approximations used in DFT
underestimate  the  value of the Coulomb  gap, while the GW
method typically yields a better value and predicts molecular
orbitals with energy ordering in better agreement with
experiments.

\begin{figure}[tbp]
\large
\begin{minipage}{0.6\columnwidth}
(a)

\includegraphics[clip,height=1\columnwidth,angle=-90]{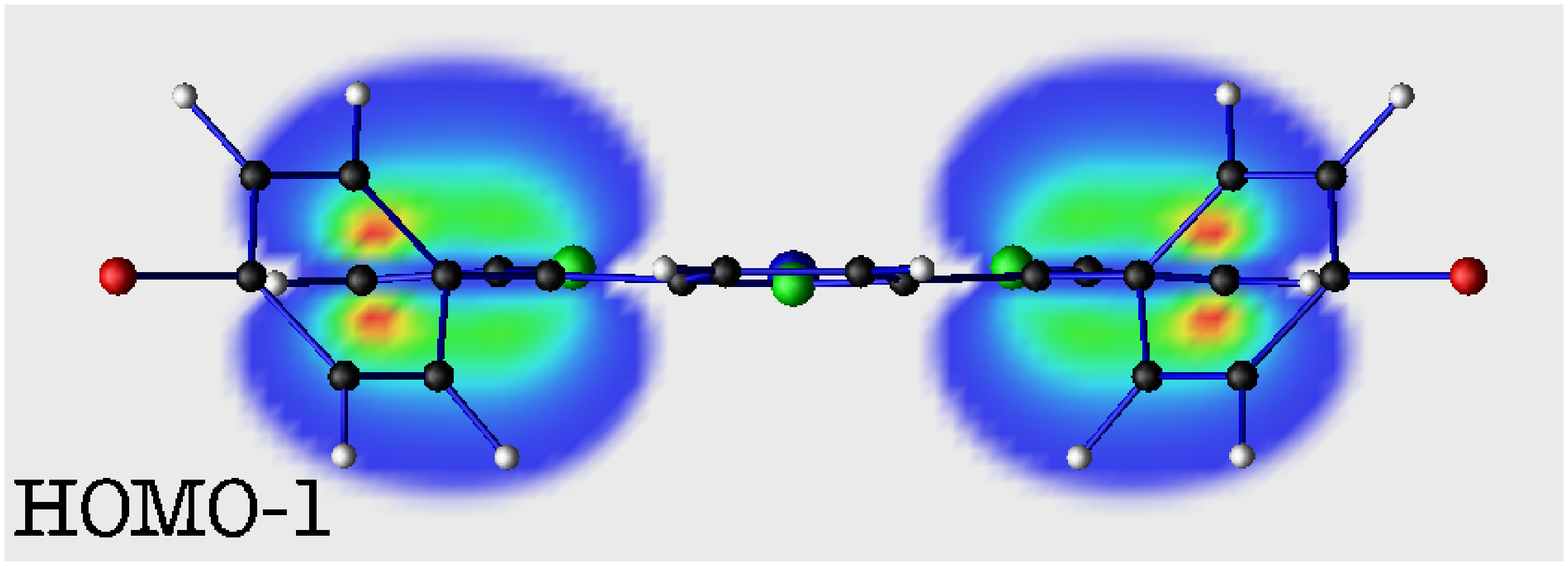}
\includegraphics[clip,height=1\columnwidth,angle=-90]{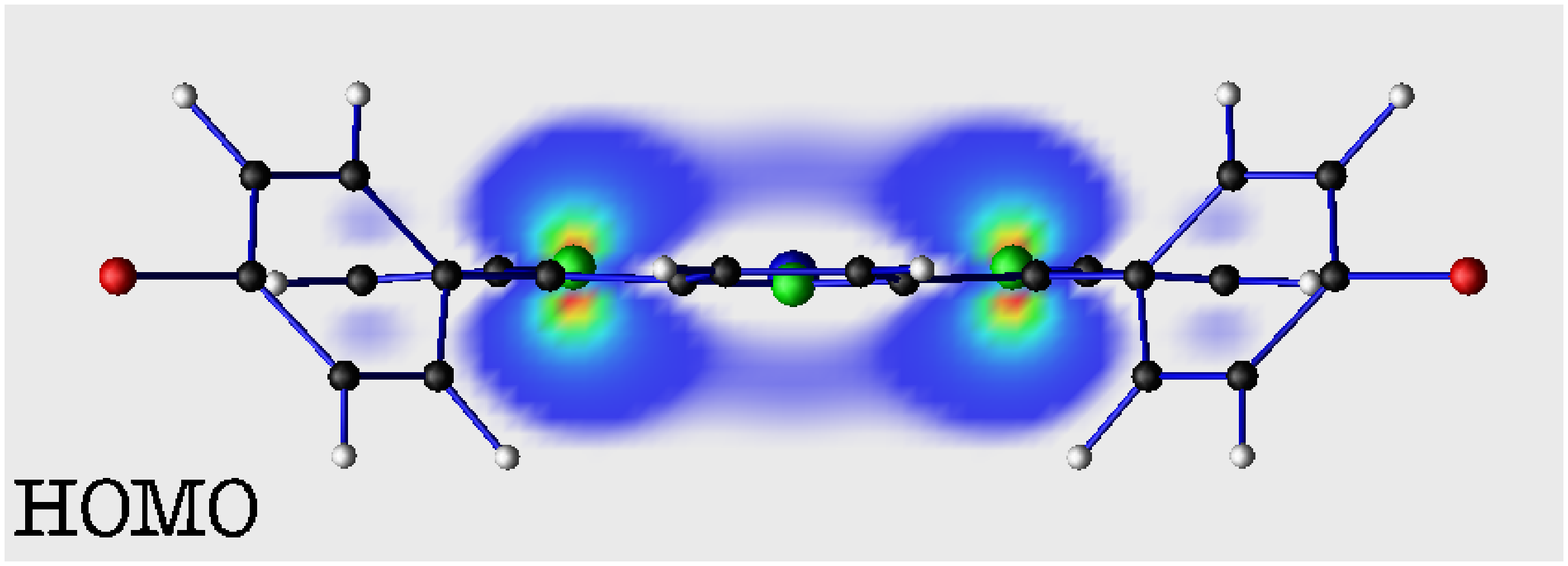}
\includegraphics[clip,height=1\columnwidth,angle=-90]{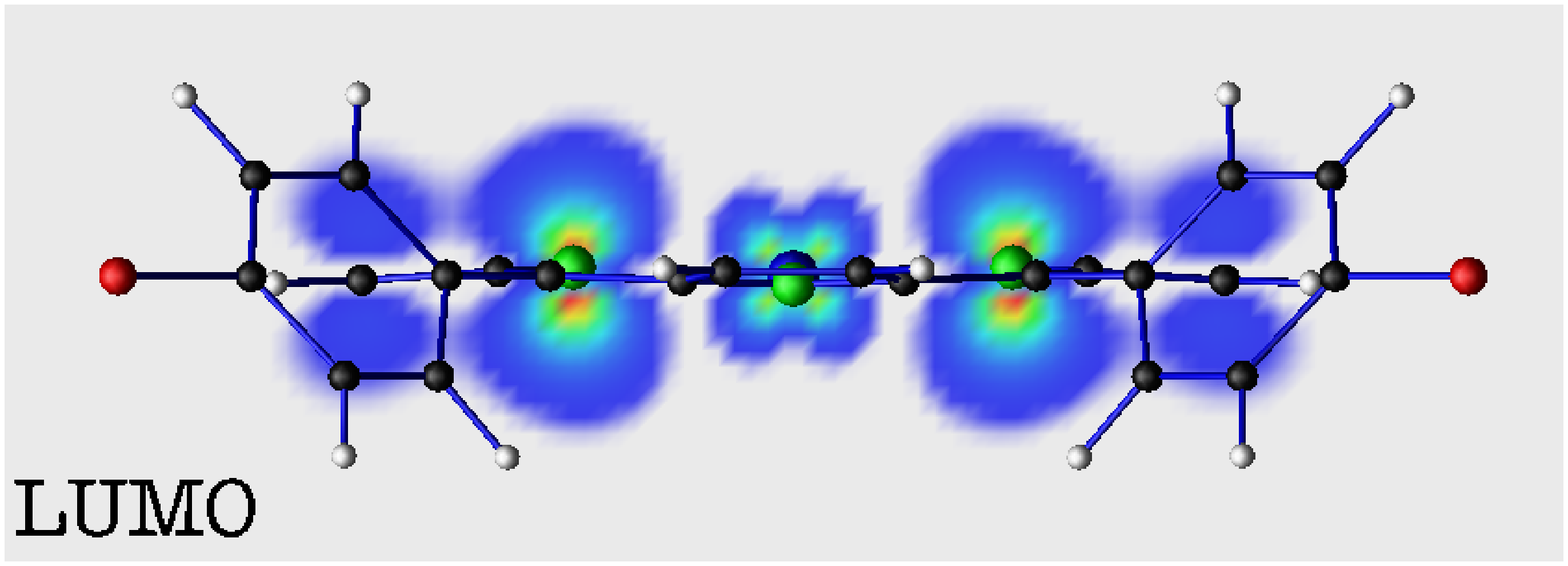}
\includegraphics[clip,height=1\columnwidth,angle=-90]{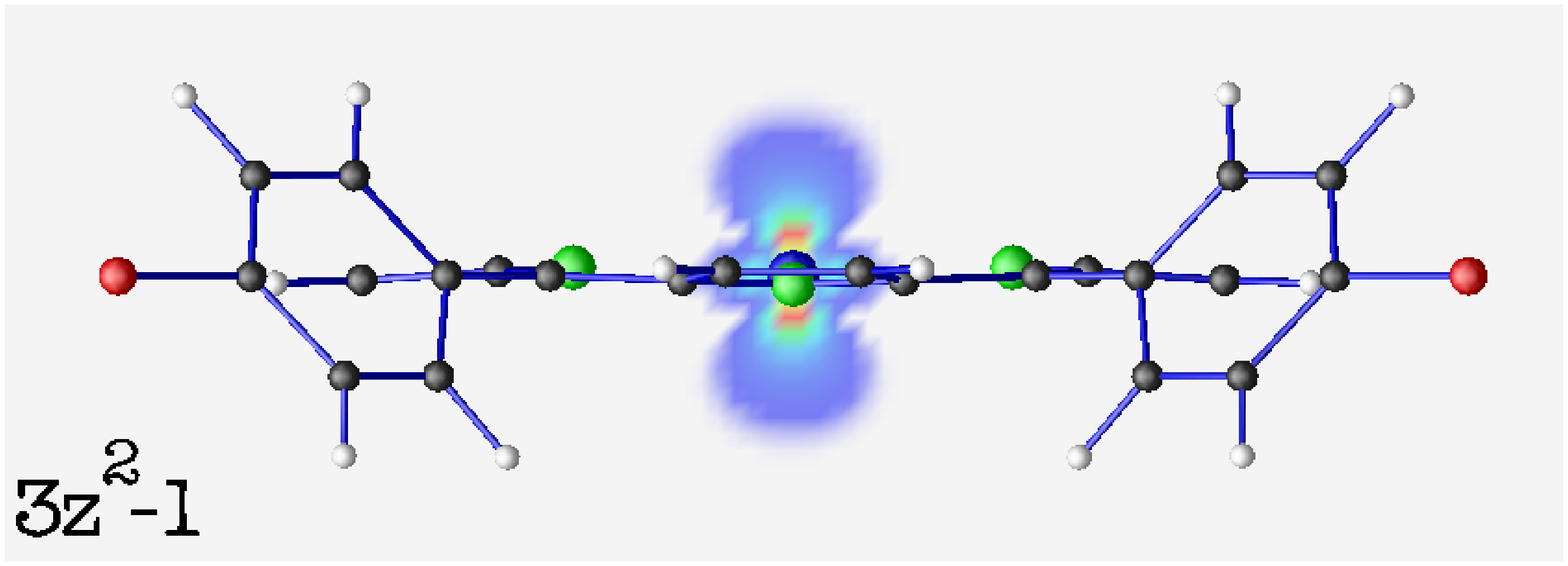}
\end{minipage}
\begin{minipage}{0.3\columnwidth}
(b)
\includegraphics[clip,width=1\columnwidth]{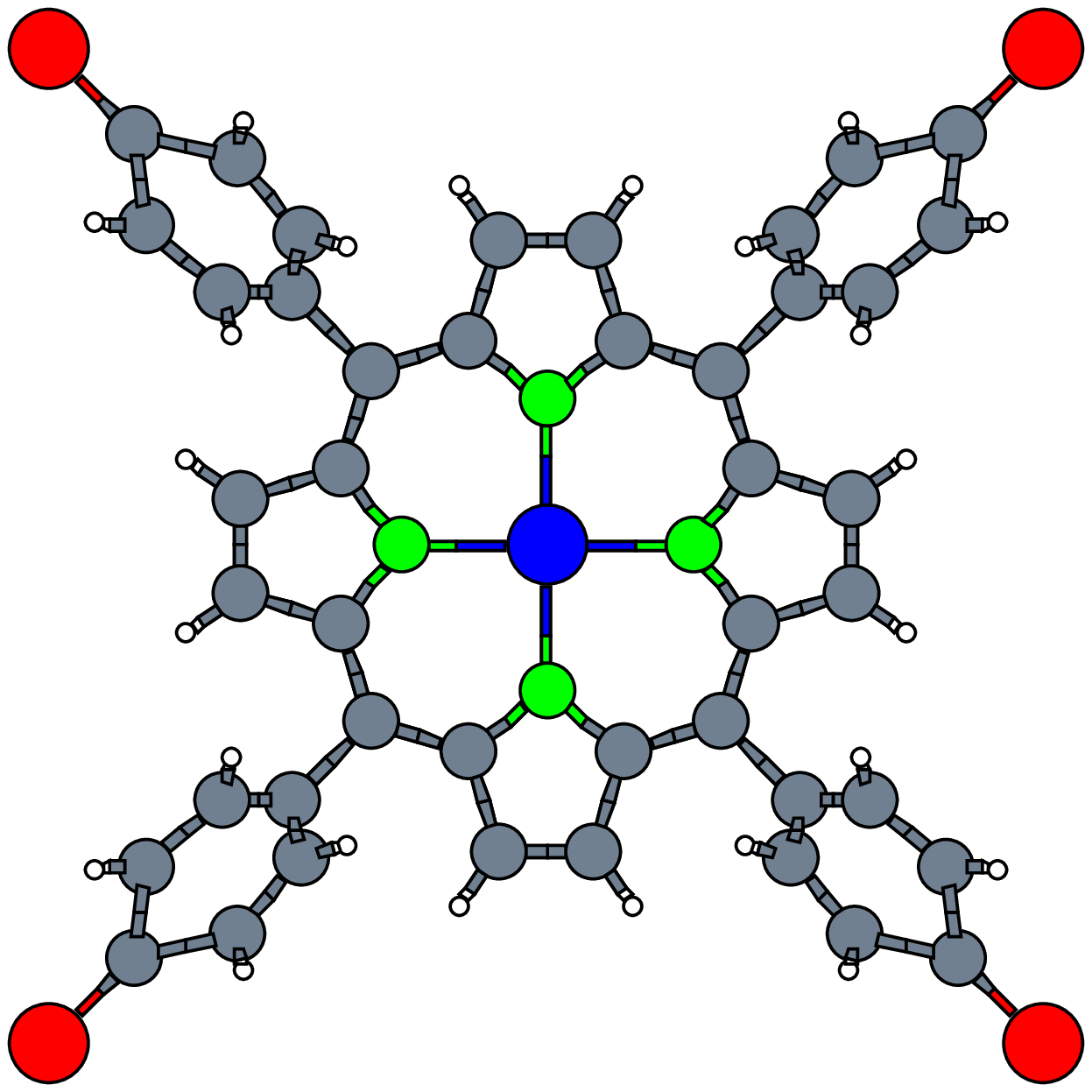}

\vspace{0.5\columnwidth} (c)
\includegraphics[clip,width=1\columnwidth]{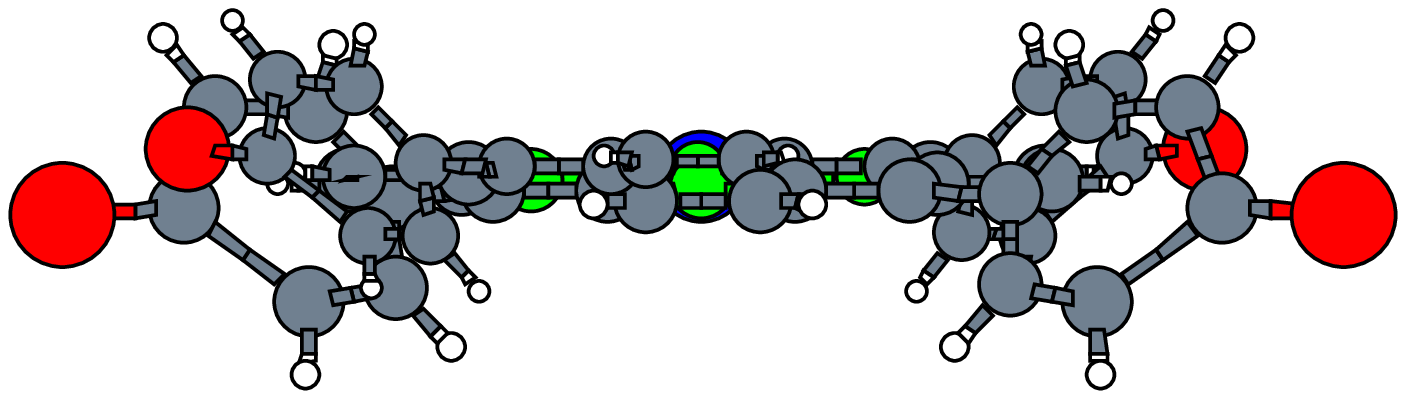}
\end{minipage}
\caption{(color online)  (a) Probability distribution of GW
molecular orbitals HOMO-1, HOMO, LUMO, and $d_{3z^2-1}$ on a
plane perpendicular to the plane of the porphyrin  ring and
crossing neighbor  atoms  N--Co--N. Hot colors (red, yellow)
indicate high  probability.  Cold colors  (green, blue) indicate
decreasing probability. The  lowest amplitude blue background was
replaced  with white,   therefore  the extent   of the dark  blue
``halo'' is somewhat arbitrary. The TBrPP-Co  molecule is shown in
(b) front view and (c) side view. Atom colors are coded as: blue
(cobalt), green (nitrogen), red (bromine), grey (carbon) and white
(hydrogen).}
\label{Fig:densities}
\end{figure}

\begin{figure}[tbp]
\includegraphics[clip,width=1\columnwidth]{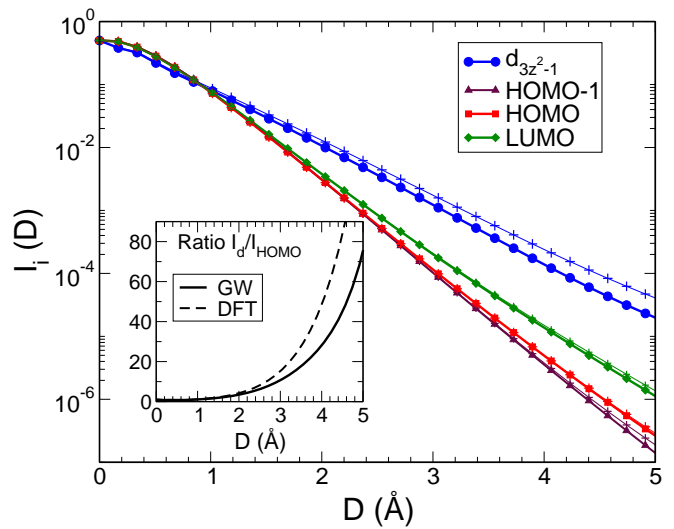}
\caption{(color online) Accumulated density of orbital $i$ (see
Eq.\
  \ref{Eq:IntD}) for orbitals $d_{3z^2-1}$,  HOMO-1, HOMO and LUMO  of
  TBrPP-Co  in gas phase. Thick  lines  with filled  symbols show  the
  accumulated density  calculated within the  GW approximation.   Thin
  lines with ``+''  signs show the corresponding accumulated densities
  calculated within DFT-GGA.  The  inset shows the ratio between integrals
  $I_d$ and $I_{\mbox{\scriptsize HOMO}}$ obtained both with GW (thick
  line) and DFT-GGA  (dashed  line). The  subsequent NRG  calculations
  were done using the GW $I_d/I_{\mbox{\scriptsize HOMO}}$ ratio.}
\label{Fig:z_int}
\end{figure}

The GW method is essentially a perturbative approach based on
calculating the electron self-energy by summing only Feynman
diagrams with low order in the screened Coulomb interaction
$W$. At lowest order, the self-energy is simply $\Sigma = i
G_0W$, which explains the name of the approximation. $G_0$ here
is the Green's function of an effective non-interacting system,
such as the Kohn-Sham (DFT) system. At the next level of
approximation, vertex corrections are added to the previous
equation: $\Sigma = i G_0 W \Gamma$. We use a simplified form
of the vertex, built from the DFT local density approximation.
More details about the construction of the LDA vertex and its
impact on the electronic structure of confined nanostructures
can be found in Refs.~\onlinecite{TiagoC06,DelsoleRG94} and
references therein. After the self-energy is calculated, we
diagonalize the quasi-particle eigenvalue equation,
\begin{equation}
\left[ { - \nabla^2 \over 2m} + V_e \right] \phi_i({\bf r}) + \int
{\rm d} {\bf r}^\prime \Sigma ({\bf r}, {\bf r}^\prime, E_i)
\phi_i ({\bf r}^\prime) = E_i \phi_i ({\bf r}) \; \; ,
\label{eq_qp}
\end{equation}
where  $V_e$ is  the  sum of  electron-ion potential  (replaced with a
pseudopotential  in   our  calculations)  and electrostatic potential created
by  electrons. Self-consistency between the Green's function and self-energy
can be added by replacing $G_0$ with a Green's function of the real system:
\cite{TiagoKHR08,KotaniSF07,Hybertsen:5390:1986} $\Sigma = i G W \Gamma$, and
repeating the calculations of self-energy, polarizability, and Green's function
until self-consistency among these quantities is found. We do not pursue full
self-consistency because it is not clear how the vertex function should be
updated during the process. Instead, we perform only one self-consistency cycle
and assume that self-consistency is usually obtained after a small number of
iterations, if the starting electronic structure is similar enough to the
converged electronic structure. \cite{KotaniSF07,ShishkinK07} The results
presented  in Figs.\ \ref{Fig:densities} and \ref{Fig:z_int} were obtained
using the $\phi_i$ orbitals (see Eq.\ (\ref{eq_qp}).

Owing to the simplicity of the subsequent Anderson model and to the fact that
spin transport is dominated by interactions in the vicinity of the central
atom, we carry out the GW calculations assuming that electronic screening in
TBrPP-Co originates predominantly from the surrounding porphyrin molecule, with
very small contribution from the central cobalt atom. Following this
assumption, we compute the screened Coulomb interaction $W$ of the non-magnetic
molecule TBrPP-Ca (with calcium in the center of the porphyrin ring) and use it
in the calculation of TBrPP-Co.

Several features of TBrPP-Co justify this methodology: (i) Cobalt $3d$ orbitals
are localized and polarize at an energy much higher than the other valence
electrons in the molecule. This is a direct consequence of spatial localization
of those orbitals and of small hybridization between Co $3d$ orbitals and
orbitals of the TBrPP compound. (ii) Co $3d$ orbitals retain most of their
identity (i.e., have a high projection on atomic Co $3d$ orbitals) even when
surrounded by the non-magnetic TBrPP structure. From that, we infer that
chemical bonds between nitrogen atoms and the central atom are weak. (iii) The
weight of $3d$ orbitals in the overall screening of TBrPP-Co is small as TBrPP
has many more valence electrons than Co. Moreover, the properties of those
electrons (energy, spatial distribution, electric susceptibility) are very
robust with respect to the type of metallic atom used, as we have checked with
DFT-GGA calculations for compounds of the series TBrPP-M (M = Mn, Fe, Co, Cu,
Ca), which show similar electronic structure and also very similar static
susceptibility.
For TBrPP-Ca, we computed the screened Coulomb interaction within the
time-dependent local density approximation (TDLDA), \cite{TiagoC06} summing
1875 virtual orbitals. The energy difference between the  highest virtual
orbital and the  LUMO is approximately 28 eV.

As Shirley and Martin reported,\cite{ShirleyM93_1} $3s$ and
$3p$ orbitals correlate strongly with $3d$ orbitals.  Thus, we
built a pseudopotential for Co  that retains   the orbitals
$3s$ and $3p$ in the  valence shell\cite{ShirleyM93_1} and
used this configuration in  all GW calculations.  In the
calculations of TBrPP-Ca,  we use a grid spacing of $0.35~a.u.$

Fig.\ \ref{Fig:levels} depicts the relative position of electronic orbitals of
gas phase TBrPP-Co, obtained  from our first-principles GW calculations.
Several orbitals  around the highest occupied  molecular orbital (HOMO)  are
strongly localized   on the  cobalt site and  are easily identified with atomic
$3d$ orbitals. All of them are occupied except  for the $d_{x^2-y^2}$,  which
is completely empty, and   the $d_{3z^2-1}$,  which  is populated by one
unpaired electron. The $z$ direction  is   perpendicular to    the  plane of
the molecule. The intrinsically molecular   orbitals (\emph{i.e.}, orbitals
with weak $3d$ character)  are non-magnetic.   The partially  occupied orbital
$d_{3z^2-1}$ gives a net spin of one half, $J^2 = 3 \hbar^2/4$, to the
molecule.  This orbital, responsible  for spin  magnetization,   is oriented
along the direction   perpendicular  to the plane    of the porphyrin ring,
giving a strong coupling with both the surface and the STM tip. These two
features confirm that  the $d_{3z^2-1}$ orbital is the origin of spin-dependent
transport in STM experiments.

We should emphasize the distinctions between the GW results presented in Fig.\
\ref{Fig:levels} and the corresponding calculation within DFT-GGA (not shown).
The most significant difference between the predictions of DFT-GGA and GW
results are in the energy difference between occupied molecular orbitals and
unoccupied molecular orbitals. This is a manifestation of the ``band-gap
underestimation", which is systematically observed when one compares these two
theories. In TBrPP-Co, DFT-GGA predicts the occupied orbitals to have energies
much higher than the GW estimate. As consequence, the HOMO-LUMO gap changes
from 2.0 eV (GW) to 0.55 eV (DFT-GGA). Additionally, the value of the
majority-minority splitting in the $d_{3z^2-1}$ orbital is $U=3.00$ eV within
DFT-GGA and $U=6.30$ eV within GW (Fig.\ \ref{Fig:levels}).\cite{CalcU}

Our  calculations indicate that,  although the $d_{3z^2-1}$
orbital is localized  around the central  atom, it would  actually
penetrate deep into the surface upon  deposition (see Fig.\
\ref{Fig:densities}). This is a consequence  of the  strong
anisotropy   of  the $d$   orbital and its orientation
perpendicular  to   the   plane of  the   molecule.   Some
non-magnetic orbitals around the  Fermi energy are also extended
along the $z$ direction.  In   fact, Fig.\ \ref{Fig:densities}(a)
shows that the HOMO-1 (second highest occupied orbital), HOMO and
LUMO extend in the out-of-plane    direction over a  similar range
as the $d_{3z^2-1}$ orbital, indicating a similar penetration
depth into the surface.

We quantify the extension of each orbital in the $z$ direction by
computing its accumulated density at a distance $D$ from a plane
containing the center of the molecule:
\begin{equation}
\label{Eq:IntD}
I_i(D) = \int_{A} d^2{\bf r} \int_{D}^{\infty} dz |\phi_i ({\bf
r},z)|^2
\end{equation}
where $i$ labels the orbital. GW  results for $I_i(D)$ for the
HOMO-1, HOMO,   LUMO   and the     $d_{3z^2-1}$ are  shown    in
Fig.\ \ref{Fig:z_int}. DFT-GGA  results  are also shown for
comparison.

The fact  that  orbitals with comparable extent and  different
angular character    exist is a strong indication     that the
high Kondo temperatures and Fano lineshapes          observed
in experiments\cite{IancuDH06_1,IancuDH06} are  the result of
scattering interference   involving   the   polarized,   localized
orbital  and non-polarized orbitals in the molecule.

As   the molecule is adsorbed   on  the surface, hybridization
between molecular and surface states will generally  lead to
charge transfer from the  surface to  the  molecule.  To verify
the  effect  of such transfer on the levels involved in the
Kondo effect, we performed DFT (PBE) calculations for the
Co-porphyrin molecule on  a monolayer of Cu(111). In these
calculations, we also used the PARSEC code but now with
periodic boundary conditions along all three spacial
directions. The monolayer was modeled as an hexagonal supercell
with 81 Cu atoms on the $xy$ plane and $26~a.u.$ separating one
monolayer from its neighboring images.

We observe  that the system relaxes  to a configuration where the
molecule-surface hybridization is stronger for levels localized in the bromine
rings, whose energies lie well below the  Fermi level. Thus, the relative
position of the energy levels involved in the Kondo effect (i.e., those close
to the Fermi energy) remains essentially unchanged once the molecule is
adsorbed. These include the cobalt-like molecular levels and the HOMO/LUMO pair
depicted in Figs.\ \ref{Fig:levels} and \ref{Fig:densities},
for which the effect of charge transfer is chiefly a uniform shift of the Fermi
energy of the system (as the chemical potentials of the molecule and the
surface are aligned), while the relative energy positions of these molecular
states and, most importantly, the net magnetization remains unchanged.

We believe such a robustness of the magnetization arises from the relatively
large separation in energy ($\gtrsim 0.5$ eV) of the Cobalt $3d$ orbitals (Fig.
\ref{Fig:levels}). The perturbation induced by the surface potential on the
cobalt atom is small in this energy scale and thus the occupation of the
orbitals does not change significantly.

\section{Effective Anderson model}
\label{sec:Model}

Using information from GW calculations, we construct a microscopic model with
realistic parameters for the system and calculate the properties in the Kondo
regime using NRG. In STM experiments with the STM tip located on top of the
Cobalt atom, the typical width of the Kondo signal corresponds to small
surface-tip bias values as compared to the characteristic spacing of electronic
levels in the molecule. In this regime, electronic transport is dominated by
elastic processes at energies close to the Fermi energy, $E_F$. This low-energy
behavior can be captured by a model that includes both the singly-occupied
$d_{3z^2-1}$ level as well as a molecular orbital $M$ with a strong projection
along the $z$ direction at the central position, and with energy
closest to $E_F$ (i.e., the HOMO in Fig.\ \ref{Fig:levels}).\cite{OtherOrbitals}

The Hamiltonian for such an Anderson-like model is written as:
\begin{equation}
\label{Eq:Hamiltonian}
H = H_{\mol}+H_{\surf}+H_{\coupling} \; .
\end{equation}
The first term describes the molecular levels involved in
transport. For the description of the Kondo regime, the geometry
of the molecule+surface+tip configuration favors orbitals with
strong $z$ projections which are either singly occupied (giving
the molecule an effective magnetic moment) or with an energy
relatively close to the Fermi level (contributing to electronic
transport). We thus have
\begin{eqnarray}
H_{\mol} & = & \sum_{\sigma} E_d \hat{n}_{d \sigma} + U \hat{n}_{d
\uparrow}\hat{n}_{d \downarrow} + \sum_{M \sigma} E_M \hat{n}_{M
\sigma} \; ,
\end{eqnarray}
with $E_i$ being the GW energy of the $d_{3z^2-1}$ orbital ($i=d$)
or some molecular orbital ($i=M$), $U$ represents the Coulomb
energy cost for double occupation in the $d$ level and $\hat{n}_{i
\sigma}$ is the occupation number of electrons with spin $\sigma$
in orbital $i$.

The remaining terms in Eq.\ (\ref{Eq:Hamiltonian}) represent the
metallic states on the surface and the molecule-surface couplings:
\begin{eqnarray}
\label{Eq:Couplings}
H_{\surf} & = & \sum_{{\bf k} \sigma} \epsilon_{{\bf k}} \hat{n}_{{\bf k} \sigma}, \nonumber\\
H_{\coupling} & = & \sum_{{\bf k}} \left( V_{d{\bf k}}
c^\dagger_{d \sigma} c_{{\bf k} \sigma} + \mbox{h. c.} \right) + \nonumber\\
& & \sum_{M,{\bf k}} \left( V_{M{\bf k}} c^\dagger_{M \sigma}
c_{{\bf k} \sigma} + \mbox{h. c.}
 \right) \; ,
\end{eqnarray}
where the operator $c^\dagger_{i \sigma}$ ($c_{i \sigma}$) creates
(destroys) and electron on orbital $i$ and $V_{i{\bf k}} \equiv
\langle \phi_i|\hat{H}| \psi_{\bf k}\rangle$ is the hybridization
matrix element between orbital $i$ and the Bloch state with vector
{\bf k} in the metal.

The effects of the noninteracting molecular level terms $(i=M)$
on the low-energy properties has been discussed in detail in Refs.\
\onlinecite{Silva:096603:2006} and \onlinecite{Dias:153304:2008} in the context
of quantum dot systems, including the possibility of a ``pseudogap Kondo
effect" for $E_M=E_F$.\cite{Silva:096603:2006,Dias:153304:2008} In the present
case, we work on a regime where $|E_M-E_F| \ll T_K$ so that the effects of the
extra level is essentially to introduce corrections to the Kondo temperature as
well as adding an extra resonance to the LDOS.

Model parameters directly obtained from first-principles
calculations include the on-site Coulomb interaction in the
$d_{3z^2-1}$ orbital ($U=6.3$~eV) and the relative energy
separation $|E_d-E_M|=1.8$~eV between the $d_{3z^2-1}$ and the
molecular orbital. The molecular-surface couplings, which give
rise to the level linewidths $\Gamma_d$ and $\Gamma_M$, can be
estimated as a function of the central atom-surface distance $D$
by combining both GW results for the orbital spatial probability
distributions  and experimental estimates from
Refs.~\onlinecite{ZhaoLCXWPWXYHZ05,IancuDH06_1,IancuDH06}, as
follows.

We consider two main contributions to the molecular level linewidths: an
intrinsic broadening $\Gamma^{\mbox{\scriptsize GW}}_{i=d,M}$ (arising from
electronic scattering processes within the molecule and obtained directly from
the GW calculations) and a contribution $\gamma_i$ from the hybridization with
both surface and bulk metallic states (described by $H_{\coupling}$ in Eq.\
(\ref{Eq:Couplings})). The key approximation involves writing $\Gamma_i \approx
\Gamma^{\mbox{\scriptsize GW}}_i + \gamma_i$, \cite{GammaGWapprox} where the
surface contribution $\gamma_i$ is directly related to the hybridization
matrix elements:
\begin{equation}
\label{Eq:gammai}
\gamma_i = -\mbox{Im } \lim_{\eta \to 0} \sum_{{\bf k}}
\frac{|V_{i{\bf k}}|^2}{E_F-\epsilon_{{\bf k}} + i \eta} \; .
\end{equation}
In general, $\gamma_i$ will depend on the details of the overlap
integrals $V_{(d,M) {\bf k}}$, particularly on the distance
between molecule and surface. While such calculations are possible
for single atoms,\cite{LinCJ05,Lin:156102:2006} they are
numerically very demanding for a molecule such as TBrPP-Co.

Instead, we take the following approach: we assume that the
relevant important metallic (bulk and surface) states leading to
the Kondo effect have long wavelengths so that the surface can be
replaced with an electron gas occupying the half-space $z
> D$. Thus, assuming the metallic states to be extended and spatially homogeneous,
$\gamma_i$ on each orbital can be related to the molecule-surface distance $D$
by taking $\gamma_i \propto I_i(D)$ given by Eq.\ (\ref{Eq:IntD}).

We estimate the proportionality constant by choosing $\Gamma_d$ to be of the
order of the d-level broadening measured in low-temperature tunneling
spectroscopy for planar TBrPP-Co on Cu(111)\cite{IancuDH06_1,IancuDH06} ($\sim
0.3$ eV) at a distance $D\sim3$\AA, taken from Ref.\
\onlinecite{ZhaoLCXWPWXYHZ05}. This sets $\Gamma_d=0.3$~eV which, together with
the calculated value for the broadening for the isolated molecule
$\Gamma^{\mbox{\scriptsize GW}}_d=0.15$ eV, gives $\gamma_d=0.15$ eV at
$D=3$\AA. From this parameter, the ratio $\gamma_d/\gamma_M$ for a given
distance $D$ can be obtained by calculating the ratio between the overlap
integrals $I_d/I_M$, given by Eq.\ (\ref{Eq:IntD}).

\section{Kondo properties}
\label{sec:Results_Kondo}

The properties of the two-orbital Anderson model given in Eq.\
(\ref{Eq:Hamiltonian}), consisting of an interacting d-level
(finite Coulomb interaction $U$) and a noninteracting molecular
orbital coupled to a metallic channel, can be obtained with NRG.
For the NRG calculations, we write the Hamiltonian given in Eq.\
(\ref{Eq:Hamiltonian}) as a single-channel Anderson
impurity model with an energy-dependent hybridization
function\cite{Silva:096603:2006,Dias:153304:2008} that includes
the information on the molecular orbitals.

We used a NRG discretization parameter $\Lambda=2.5$ and kept
up to 1000 states (not including SU(2) spin degeneracy) on each
iteration; we use a wide-band approximation, which assumes
$\Gamma_d \ll D_{\mbox{\scriptsize band}}$, where $D_{\mbox{\scriptsize band}}$
is the bandwidth of the continuum of metallic states (which includes
contributions from both surface and bulk states). To check this assumption, we
performed DFT calculations for bulk Cu, which give
$D_{\mbox{\scriptsize band}} \approx 10$ eV. The molecular  density of states
(DOS), given by ($-\pi^{-1}$)Im$[G_d(\omega)+G_M(\omega)]$,  was obtained from
the interacting Green's functions $G_{d(M})(\omega)\equiv \langle \langle
c_{d(M)}:c^{\dagger}_{d(M)} \rangle \rangle_{\omega}$ calculated from NRG
spectra.\cite{Silva:096603:2006,Dias:153304:2008} The DOS calculated in such
manner contains spectral information from both ``d" and ``M" levels, which,
based on their wave function $z$-extension (Fig.\ \ref{Fig:z_int}), should
yield comparable contributions to the STM signal.

In the calculations, the orbital energies $E_d$ and $E_M$
entering the model are referenced to the Fermi energy $E_F$ at
the metallic surface. While DFT gives a crude estimate for $E_F$ (Fig.\ \ref{Fig:levels}), we should
stress it is numerically challenging to extract a more reliable estimate for $E_F$ from a
corresponding GW calculation. Instead, we chose to treat $E_d-E_F$ as a free parameter.
Thus, the surface-molecule charge transfer, identified in the DFT calculations as the
mechanism controlling the relative position of the molecular orbitals to the
Fermi energy, can be studied by changing $E_d$ and $E_M$ while keeping
$E_d-E_M$ constant, which is equivalent to effectively varying the chemical
potential in the molecule.

\begin{figure}[tbp]
\includegraphics[clip,width=1\columnwidth]{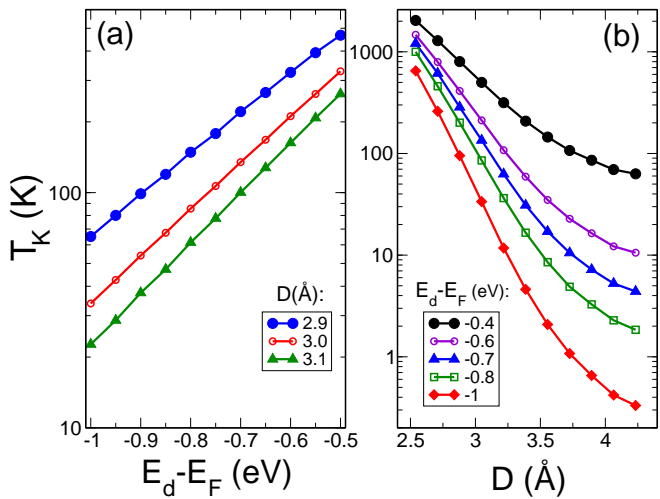}
\caption{(color online) Kondo temperature $T_K$ (obtained from NRG
calculations with GW-based parameters) versus $E_d-E_F$ (a) and
molecule-surface distance $D$ (b). }
\label{Fig:TK_D_mol}
\end{figure}

Figure \ref{Fig:TK_D_mol} shows the Kondo temperature $T_K$ as
a function of $E_d-E_F$   (Fig.\ \ref{Fig:TK_D_mol}a)   and of
the central atom-surface  distance  $D$ (Fig.\
\ref{Fig:TK_D_mol}b), for different $E_d-E_M$ values, as shown.
We find that the calculated Kondo temperatures are within the
range of experimentally observed Kondo temperatures in similar
systems \cite{ZhaoLCXWPWXYHZ05,IancuDH06_1,IancuDH06} for $D
\approx 3$~\AA, and charging $-0.8 \leq E_d-E_F \leq -0.6$~eV
(note that, in this regime, orbital $M$ becomes the LUMO). For
the value $E_d-E_F= -0.7$~eV reported in
Refs.~\onlinecite{IancuDH06_1,IancuDH06}, we obtain $T_K \sim
140$~K for $D \approx 3$\AA, in the same range as the reported
experimental results.

Overall, we  obtain an exponential dependence of $T_K$ with
both $D$  and $E_d-E_F$. Such  behavior of $T_K$ is not
surprising:  for $\Gamma_M  \ll \Gamma_d$,  the many-body
Hamiltonian reduces  to the usual single impurity Anderson
model  and one expects $T_K$ to  follow an exponential
behavior:\cite{HewsonBook}
\begin{equation}
T_K \sim \sqrt{\frac{U \Gamma_d}{2}} e^{-\pi|E_d-E_F||E_d-E_F+U|/2U\Gamma_d} \; .
\end{equation}
The rapid decrease of $T_K$ with $D$ follows naturally from the
dependence of the accumulated density $I_d$ (and hence $\Gamma_d$)
on $D$, as depicted in Fig.\ \ref{Fig:densities}.

Interestingly,   both    exponential  behaviors  can,    in   a
sense, compensate  each other if a  conformational change in the
molecule decreases $D$  and  $E_d$ at the    same time, resulting
in  a weaker variation in $T_K$.  In fact,  this explanation
is consistent with the behavior reported in
Ref.~\onlinecite{IancuDH06_1}, for   which the structural change
from saddle to planar conformation (reducing  $D$) was accompanied
by  a  slight reduction of $0.2$~eV in $E_d-E_F$, leading to a
moderate increase  in the Kondo temperature of $\Delta T_K/T_K
\sim 0.3$.

\begin{figure}[tbp]
\includegraphics[clip,width=1\columnwidth]{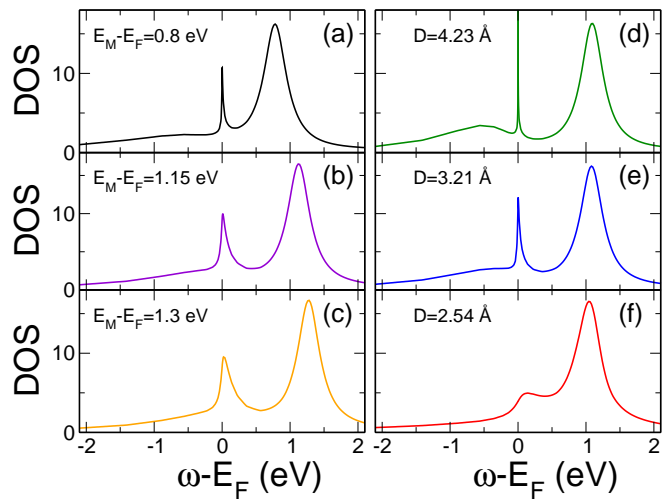}
\caption{(color online) Molecular density of states (DOS) system
for different molecular  charging (a-c,   $D=3$ \AA)  and
molecule-surface distances (d-f, $E_M-E_F=1.1$ eV). A Kondo
resonance at the Fermi energy and the HOMO and LUMO  orbitals are
shown. The width variation of the  Kondo resonance ($\sim T_K$) is
consistent with the results in Fig.\ \ref{Fig:TK_D_mol}.}
\label{Fig:DOS}
\end{figure}

The width of the Kondo resonance  in the energy-resolved
density of states (DOS) of the adsorbed molecule is
proportional to the Kondo temperature.  Experimentally, the
molecule DOS can be directly probed with STM  by suppressing
the   direct tunneling  from    tip   to
surface.\cite{OtteTBLBLHH08} We have calculated the DOS from
the Green's  functions  for the  model with  NRG. Figure
\ref{Fig:DOS}  depicts   the  DOS  for  different charge
transfer ($E_M-E_F$) and distance ($D$) values.  In all cases,
the molecular LUMO resonance is prominent, as well as  the
Kondo peak at the Fermi energy.  The $E_d$ level is present but
it  is not as prominent. The  width of the  Kondo peak
increases substantially for smaller  distances  and changes
with charging,  correlating well  with    the    behavior  of
$T_K$   in Fig.\ \ref{Fig:TK_D_mol}. Notice that the  Kondo
feature  can be extremely sharp (a  couple of millivolts
across)   for larger molecule-surface distances (Fig.\
\ref{Fig:DOS}d).

\section{Summary}
\label{sec:Summary}

In summary, we   combine first principles GW  calculations with
the numerical  renormalization-group  method  and  calculate
Kondo temperatures of TBrPP-Co, a magnetized porphyrin
compound, adsorbed on copper. This allows for a quantitative
comparison with experimental results, provided  that two
crucial parameters are  known: the separation between molecule
and surface atoms, and the amount of charge transfer to or from
the molecule  (or, alternatively, the energy of  the HOMO
relative to the Fermi    energy of the surface). We find that
our \textit{ab initio} treatment beyond DFT is essential, as
the contribution from dynamical correlations to the molecule
parameters (e.g.,  $U$, orbital energies  and hybridizations)
are important.

Our first-principles GW calculations indicate that there are
two kinds of orbitals in the vicinity of the Fermi energy with a strong overlap
with the surface: a spin-polarized $d_{3z^2-1}$ orbital, which gives  rise to
the  Kondo effect, and non-polarized molecular orbitals. Both types have strong
vertical projections and should dominate the observed STM conductance signal.
Numerical renormalization-group calculations for a microscopic model built
based on these GW results yield Kondo temperatures in good agreement to those
observed in recent STM measurements.\cite{IancuDH06_1,IancuDH06} Additionally,
the calculated DOS shows both Kondo and molecular resonances consistent with
the tunneling spectroscopy data.

We find, not unexpectedly, that the Kondo temperature and the DOS features
depend strongly on the  interplay between two parameters: the position  of the
partially occupied spin-polarized orbital with respect to the Fermi energy,
which increases $T_K$ as it approaches  that energy, and    the wave function
overlap between molecular orbitals and surface states, which decreases  $T_K$
as it reduces (increasing $D$). Most interestingly, in a regime where these two
parameters compensate each other, $T_K$  may show non-exponential behavior
providing a natural explanation to recent STM measurements of $T_K$.

We should note the combination of numerical methods employed
here is quite general and can be used to study other metallo-organic molecules.
Examples are porphyrin molecules with different magnetic atoms in the center
such as Fe, Cu and Mn. Although the low-energy model extracted from the GW
calculations in such cases will, in general, involve multiple orbitals and
higher spin configurations, such situations can be accommodated within
multi-orbital extensions of the NRG method,\cite{BullaCP08} bringing the
interesting prospect of a comprehensive theoretical description of such
strongly correlated molecular systems.

\acknowledgments

We acknowledge enlightening discussions with Violeta Iancu, Saw
Hla, Nancy Sandler, Kevin Ingersent, Enrique Anda and Enrique
Louis. Research performed at the Materials Science and
Technology Division, sponsored by the Division of Materials
Sciences Engineering BES, U.S. DOE under contract with
UT-Battelle, LLC. Computational support was provided by the
National Energy Research Scientific Computing Center (NERSC).
SEU acknowledges support from NSF grants DMR-0336431, 0304314
and 0710581. LGGVDS and ED acknowledge support from NSF grant
DMR-0706020.


\begin{thebibliography}{44}
\bibitem{CalcU} As a general rule, we can use the majority-minority
    splitting from GW to estimate $U$ for partially  occupied  $d$-orbitals in
    other TBrPP-M (M = transition metal atom) compounds.
    The strength of $U$  depends crucially on the total magnetization of the
    molecule (TBrPP-Mn, for instance, has values of $U$ several eVs larger than TBrPP-Co)
    and on the strength of the TBrPP molecular field.
\bibitem{OtherOrbitals} The HOMO-1, HOMO-2 and HOMO-3 orbitals, although closer
    in energy to $E_F$ than $3d_{z^2}$, are localized further away from the center of
    the molecule level, with a high amplitude on the phenyl-bromine radicals.
    Thus, their contribution to the STM signal with the tip placed on top of
    the Co atom is negligible.
\bibitem{GammaGWapprox} In this approximation, we are neglecting many-body
contributions due to the Coulomb interactions between the electrons in orbital
$i$ entering the model (e.g., Cobalt $d_{3z^2-1}$ electrons) and the electrons
in the surface. Such contributions are likely to be much smaller than the
interaction effects of electrons within orbital $i$, which are well captured
by the GW calculation in vacuum.
\end{thebibliography}

\end{document}